\documentclass[manuscript,screen,acmlarge]{acmart}

\setcopyright{none}
\renewcommand\footnotetextcopyrightpermission[1]{} %
\AtBeginDocument{%
  \providecommand\BibTeX{{%
    \normalfont B\kern-0.5em{\scshape i\kern-0.25em b}\kern-0.8em\TeX}}}
    
\usepackage{graphicx}
\graphicspath{{figures/}}
\usepackage[utf8]{inputenc}
\usepackage{siunitx}
\usepackage{wrapfig}
\usepackage{soul}
\usepackage{multirow}

\DeclareMathAlphabet{\mathcal}{OMS}{cmsy}{m}{n}

\def\mysection#1#2{\section{#1}\label{sec:#2}}

\def\mysubsection#1#2{\subsection{#1}\label{sec:#2}}

\newcommand{\refSec}[1]{Section~\ref{sec:#1}}
\newcommand{\refFig}[1]{Figure~\ref{fig:#1}}

\newcommand{\refTbl}[1]{Table~\ref{tbl:#1}}

\newcommand{\etal}{\textit{et al}. }

\DeclareSIUnit\cpd{cpd}
\DeclareSIUnit\ppi{ppi}
\DeclareSIUnit\FPS{FPS}
\DeclareSIUnit[number-unit-product = ]\inch{\char`"}
\DeclareSIUnit\pix{px}
\DeclareSIUnit\pixel{px}
\DeclareSIUnit\Hz{Hz}

\newcommand{\SIrangeto}[3]{\SIrange[range-phrase={ to }]{#1}{#2}{#3}}

\newcommand{\SIrangethru}[3]{#1-\SI{#2}{#3}}

\newcommand{\SIrangebys}[3]{#1$\times$\SI{#2}{#3}}

\NewDocumentCommand\angRange{O{} m m}{\SIrange[parse-numbers=false, #1]{\ang[parse-numbers=true]{#2}}{\ang[parse-numbers=true]{#3}}{}}

\begin{document}

\title{Learning GAN-based Foveated Reconstruction to Recover Perceptually Important Image Features}

\author{Luca Surace}
\email{luca.surace@usi.ch}
\affiliation{
  \institution{Università della Svizzera italiana}
  \country{Switzerland}
}

\author{Marek Wernikowski}
\email{marek.wernikowski@zut.edu.pl}
\affiliation{%
  \institution{Università della Svizzera italiana}
  \country{Switzerland}
}
\affiliation{%
  \institution{West Pomeranian University of Technology}
  \country{Poland}
}

\author{Cara Tursun}
\email{cara.tursun@rug.nl}
\affiliation{%
  \institution{Università della Svizzera italiana}
  \country{Switzerland}
}
\affiliation{%
  \institution{University of Groningen}
  \country{Netherlands}
}

\author{Karol Myszkowski}
\email{karol@mpi-inf.mpg.de}
\affiliation{%
  \institution{Max Planck Institute for Informatics}
  \country{Germany}
}

\author{Radosław Mantiuk}
\email{rmantiuk@wi.zut.edu.pl}
\affiliation{%
  \institution{West Pomeranian University of Technology}
  \country{Poland}
}

\author{Piotr Didyk}
\email{piotr.didyk@usi.ch}
\affiliation{%
  \institution{Università della Svizzera italiana}
  \country{Switzerland}
}

\begin{abstract}
A foveated image can be entirely reconstructed from a sparse set of samples distributed according to the retinal sensitivity of the human visual system, which rapidly decreases with increasing eccentricity. The use of Generative Adversarial Networks has recently been shown to be a promising solution for such a task, as they can successfully hallucinate missing image information. As in the case of other supervised learning approaches, the definition of the loss function and the training strategy heavily influence the quality of the output. In this work,we consider the problem of efficiently guiding the training of foveated reconstruction techniques such that they are more aware of the capabilities and limitations of the human visual system, and thus can reconstruct visually important image features. Our primary goal is to make the training procedure less sensitive to distortions that humans cannot detect and focus on penalizing perceptually important artifacts. Given the nature of GAN-based solutions, we focus on the sensitivity of human vision to hallucination in case of input samples with different densities. We propose psychophysical experiments, a dataset, and a procedure for training foveated image reconstruction. The proposed strategy renders the generator network flexible by penalizing only perceptually important deviations in the output. As a result, the method emphasized the recovery of perceptually important image features. We evaluated our strategy and compared it with alternative solutions by using a newly trained objective metric, a recent foveated video quality metric, and user experiments. Our evaluations revealed significant improvements in the perceived image reconstruction quality compared with the standard GAN-based training approach.

\end{abstract}

\maketitle

\begin{wrapfigure}{R}{0.5\textwidth}
    \centering
    \includegraphics[width=0.5\textwidth]{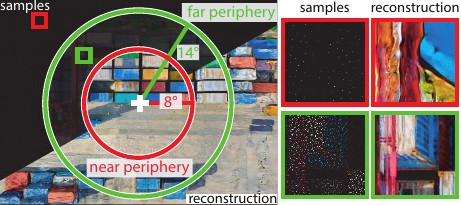}
    \caption{The input to the foveated image reconstruction are sparse samples (magnify the top-left part of the image or see the insets), based on which the technique reconstructs the images (bottom-right part of the image). The image in this example consists of three regions: fovea (100\% samples), near periphery (12\% samples), and far periphery (0.7\% samples). The white cross in the center indicates the position of the gaze. The reconstructed regions are combined by using linear blending to create a smooth transition between them.
    }
    \label{fig:samples_visualization}
\end{wrapfigure}

\mysection{Introduction}{introduction}
Wide-field-of-view displays, such as virtual and augmented reality headsets, require efficient methods to generate and transmit high-resolution images. Techniques for reconstructing foveated images seek to solve the problem by leveraging the non-uniform sensitivity of human vision to spatial distortions across a wide field of view and generate high-quality images around only the location of the gaze as indicated by an eye-tracking device. Such foveated systems usually consist of two main steps \cite{deepfovea,stengel_adaptive_2016,Tursun2019}. First, an image is generated or transmitted in the form of a sparse set of samples that are generated according to the location of the gaze. Second, the image is reconstructed from the sparse information before being shown to the observer. An example of such a technique is foveated rendering \cite{guenter_foveated_2012} where fewer image samples are computed for peripheral vision to save computation during rendering (\refFig{samples_visualization}).

We focus on the second step above, i.e., reconstructing an image from sparse samples. While simple techniques, such as interpolation \cite{stengel_adaptive_2016} can be used to this end, it has been demonstrated that machine-learning techniques, more precisely, generative adversarial networks (GANs), can provide superior results \cite{deepfovea} due to their ability to hallucinate missing content based on the learned statistics of the given image or video. Although such reconstruction requires additional computation, it can provide results of a similar quality as those obtained by simpler techniques while using fewer input samples. However, several challenges persist in design and training in this regard. Every GAN architecture is composed of two neural networks trained simultaneously \cite{Goodfellow2014,deepfovea}. In the task of foveated image reconstruction, one of them, called the \emph{generator}, is responsible for reconstructing the image from a sparse set of samples, while the second one, called the \emph{discriminator}, is responsible for discriminating between real and reconstructed images. The training iterations of both networks are interleaved such that an improvement in one of them triggers an improvement in the other. Ultimately, training in this manner can be viewed as a game between the generator and the discriminator networks, where the generator tries to reconstruct perfect images from a limited number of samples to try to fool the constantly improving discriminator. Therefore, to successfully train a GAN architecture, maintaining a balance between the training of the generator and the discriminator is highly important. Another challenge, which is the main focus of this paper, involves the choice of training loss and procedure. As in other supervised learning solutions, they have a significant influence on the final performance of GAN-based reconstruction.

The recent literature has acknowledged that for any task where the perceived quality is critical, the loss function must capitalize on visually important image features. A well-known strategy to incorporate such perceptual findings in neural network training is to use a perceptual loss function (e.g., LPIPS \cite{zhang2018unreasonable}). While training a GAN, it is possible to use such a function as the loss for training the generator \cite{deepfovea}. However, our main hypothesis is that this is insufficient because the training of the discriminator should alsotake into account the properties of human visual perception. To address this problem, we propose a new training scheme for the discriminator. Instead of training the discriminator to distinguish reconstructed images from real images, our technique trains it to distinguish the reconstructed images from real images that contain imperceptible distortions.

In this way, the discriminator network can inherit the limitations of the HVS (Human Visual System) represented in the data, and stops penalizing the generator network for reconstructing imperfect images within perceptual limits.

Consequently, in this work, we aim to improve GAN-based machine-learning approaches for foveated reconstruction by introducing a new training scheme based on perceived quality degradation. We first design and conduct psychophysical experiments to study the sensitivity of human visual system to content-based hallucinations across a wide visual field. Our choice of stimuli is inspired by several findings on the degradation of the sensitivity of human vision along the periphery. Although many foveated systems have previously exploited the loss of visual sensitivity to high-frequency information \cite{guenter_foveated_2012,Tursun2019}, the effect does not fully explain the visibility of missing information in peripheral vision. For example, past psychophysical experiments suggest that even though a perfect reconstruction of fine details of the image in the periphery is not critical, a complete lack of high spatial frequencies is detectable \cite{thibos1987}. Another effect that is not fully explained by the reduced visual sensitivity to higher spatial frequencies in the periphery is the increased positional uncertainty \cite{levi1987, hess1993increased}. To reflect these findings, we employ a technique of texture synthesis guided by the statistics of the original images to generate stimuli with varying amounts of hallucinated content. We argue that this type of distortion resembles the content synthesized by using GAN-based image reconstruction, and our experiments quantify their visibility. We then demonstrate how to incorporate the experimental results into training. Finally, we show how our strategy of focusing on perceptually important image features during training can lead to a GAN-based foveated reconstruction method that provides higher reconstruction quality with the same number of input samples or, conversely, the same perceived quality using fewer samples, leading to savings in bandwidth or rendering time. We argue that this is possible because our foveated reconstruction method aims to recover perceptually important image features that would be otherwise lost due to sub-sampling. The new dataset also allows us to calibrate application-specific objective metrics that predict image quality. We use the new metric and the perceptual experiments to evaluate our new training strategy and compare it with alternative solutions.

\mysection{Related work}{related_work}
Our work takes inspiration from and bridges the expertise in visual perception, computer graphics, and machine learning. Here, we provide an overview of the relevant works from these fields.

\mysubsection{Foveal vs. peripheral vision}{foveal_vs_peripheral_vision}

\paragraph{Retina} The perceptual capabilities of the HVS have been extensively studied under different positions of visual stimuli in the visual field. Perception is not uniform across the visual field owing to optical and physiological limitations. Studies on the retina revealed that the density of photoreceptors in the retina is highly heterogeneous \cite{curcio1990human, watson2014formula}. The central region of the retina, called the \emph{fovea centralis} (or \emph{fovea}), is characterized by a relatively high density of cone photoreceptors and retinal ganglion cells (RGCs). This provides foveal vision with a superior perceptual capability compared with non-foveal (or \emph{peripheral}) vision. Although the fovea provides a sharp central vision, it is relatively small and corresponds to approximately \SI{2}{\degree} of the visual field, which spans up to \SIrangethru{160}{170}{\degree} \cite{koob2010}. On the contrary, peripheral vision corresponds to more than \SI{99}{\percent} of our visual field.

\paragraph{Peripheral contrast sensitivity} To study the differences between foveal and peripheral vision, previous psychophysical studies have focused on measurements of the contrast sensitivity function (CSF), which represents the sensitivity to changes in contrast at different spatial frequencies \cite{mannos1974, barten1999, kim2013}. Research on the fovea has shown that the human CSF curve has a peak around \SIrangethru{4}{8}{cycles\,per\, degree\,(cpd)}, with its tail reaching up to \SIrangethru{50}{60}{\cpd}. Later, Peli \etal \cite{peli1991image} and, more recently, Chwesiuk \etal \cite{chwesiuk2019csf} extended these measurements to peripheral vision, and observed that the decline in contrast sensitivity is characterized by a smaller peak that shifts toward lower spatial frequencies as eccentricity increases. This implies a loss of sensitivity to content with high spatial frequency content in peripheral vision.

\paragraph{Foveated rendering} The differences between foveal and peripheral vision mentioned above have led to gaze-contingent techniques that process and display images depending on the position of the gaze of the observer. Foveated rendering is an actively studied gaze-contingent technique in this domain. It uses the position of the gaze from an eye tracker for a low-resolution image reconstruction in the periphery  \cite{guenter_foveated_2012, patney_2016, stengel_adaptive_2016, meng_kernel_2018, kim_foveated_2019, bruder_voronoi-based_2019}. These studies have significantly reduced the computational cost of rendering because they reduce the number of pixels to be rendered \cite{weier2017perception}. However, their reconstruction methods are mostly based on the simple interpolation of a sub-sampled image and such post-processing steps as temporal antialiasing and contrast enhancement. However, such a simple reconstruction approach does not aim to replace the high-frequency spatial details lost as a result of the undersampling of the underlying content, leading to noticeable degradation in quality.

\paragraph{Hallucinating image details} Psychophysical measurements show that peripheral vision requires a more sophisticated model than a simple boundary between perceptible and imperceptible regions of contrast guided by the shape of the CSF \cite{rosenholtz_summary_2012, rosenholtz_capabilities_2016}. Thibos \etal \cite{thibos1987} revealed that the threshold of resolution declines from \SIrangeto{14}{2.6}{\cpd} in the range of eccentricity of \SIrangeto{5}{35}{\degree}, whereas the threshold of detection drops from \SIrangeto{46}{28}{\cpd} in the same range of eccentricity. As a result of the faster drop-off in the threshold of resolution, there exists a band of spatial frequencies that can be detected but not accurately resolved for each value of eccentricity. Additional studies have shown that performance in terms of discriminating the spatial phase also degrades with increasing eccentricity and leads to greater positional uncertainty in visual perception \cite{rentschler1985, morrone1989, levi1987}. Rosenholtz \cite{rosenholtz2011} claimed that the HVS encodes image statistics rather than precise location information in peripheral vision, leading to a performance decline in resolving the stimulus position. These studies have important implications for the design of foveated image reconstruction methods because they clearly show that HVS models driving the reconstruction must be comprehensive enough to consider multiple aspects of visual perception. In contrast to the standard reconstruction techniques mentioned above, we address this missing piece in the foveated image reconstruction pipeline.

\mysubsection{Metamers}{metamers}
The goal of foveated rendering can be viewed as the low-cost production of images that are metameric to the full-quality rendering. \emph{Metamer} here refers to images that are structurally different but appear the same to the human observer. Moreover, foveated rendering assumes knowledge about gaze position; therefore, images are metameric usually only for a given gaze location at which most of their content is observed by peripheral vision.

The limitations of the HVS perception have inspired several important studies on metamerism. Initial work aimed to synthesize textural metamers, i.e., different images representing the same type of texture. To this end, Portilla and Simoncelli \cite{simoncelli2000} used an iterative optimization that is run until a randomly initialized image patch converges to the same summary statistics as the target texture. Their observations led to further studies on crowding effects, and Balas \etal \cite{balas2009} revealed that the representation of summary statistics can explain the crowding effects observed in the periphery. Rosenholtz \etal \cite{rosenholtz_summary_2012, rosenholtz2012rethinking}  introduced the texture tiling model (TTM), which models the performance of visual search in the periphery. Based on the ideas on summary statistics, Fridman \etal \cite{fridman2017ffcn} proposed a convolutional network to reproduce the outputs of the TTM, and Deza \etal \cite{deza_towards_2017} introduced a generative adversarial network model for creating metamers of peripherally viewed natural images. Instead of using hand-crafted summary statistics, as previously introduced by Portilla and Simoncelli \cite{simoncelli2000}, they used channel autocorrelation statistics computed from the pre-trained VGG network features \cite{vgg}. Although these studies have delivered promising results, their main goal is to study foveal texture perception or provide a reference model for studying the properties of peripheral vision (e.g., for visual crowding). More recently, Walton \etal \cite{walton2021beyond} proposed a real-time method for producing metameric images to peripherally viewed input images with the main application of image and video compression.

While all the above-mentioned studies have focused on producing a metamer of an input image, this approach is not directly applicable to foveated rendering, the goal of which is to avoid rendering the original, full-quality image in the first place. Therefore, it is interesting to consider the problem of computing images that are metameric to full-quality rendering but are derived based on partial information from the rendering system. Examples of such techniques include standard foveated rendering, where the shading rate is reduced toward the periphery \cite{guenter_foveated_2012}, as well as more recent techniques, where contrast enhancement \cite{patney_2016} or noise synthesis \cite{tariq2022noise} are applied to further reduce the amount of information needed for reconstructing peripherally viewed metamers of full-quality images. Kaplanyan \etal \cite{deepfovea} recently introduced a powerful method to this end. They proposed a foveated image compression solution by using a GAN model that reconstructs perceptually plausible image sequences from a very sparse set of samples while maintaining temporal coherence. Our work takes inspiration from this solution and aims to minimize the required number of samples from the underlying content while achieving the best-perceived quality. To this end, similar to past work, we focus on reconstructing by hallucination but capitalize on positional uncertainty, and distortions \cite{thibos1987} by modifying the training scheme. More precisely, in contrast to the scheme presented by Kaplanyan \etal \cite{deepfovea}, where the discriminator network is trained on ground-truth images, we train it on data derived from a series of systematic experiments that analyze the sensitivity of the HVS to distortions in the periphery.

\mysubsection{Image metrics and perceptual loss}{foveated_quality_metrics}

One way of guiding image reconstruction is to use image metrics. Current foveal quality metrics use the properties of central vision and provide inaccurate predictions for the periphery. The growing research on peripheral vision and its applications to foveated image reconstruction suggests the need for new foveated metrics \cite{wang_foveated_2001, sanghoon_lee_foveated_2002, rimac_foveated_2010, tsai_foveation_2014, swafford_user_2016, hsu_is_2017, vranjes_foveation_2018, guo_perceptual_2018, tran_impacts_2019}. These metrics are promising candidates to guide the loss function in learning-based approaches. However, their complex implementations, costly computations, and, in some cases, non-differentiable operations pose challenges for training models of image reconstruction by using them. An alternative to this is to use a training loss defined on the feature maps from a pre-trained deep network. This has become one of the most common approaches to learning-based image reconstruction, especially for super-resolution-based techniques \cite{johnson_perceptual_2016, dosovitskiy2016, zhang2018unreasonable}. Compared with a simpler loss function, such as the mean squared error (MSE), the loss functions defined on the hierarchical features of deep networks more closely resemble how the HVS processes visual information. However, there may still be significant differences between deep network representation and human visual perception \cite{feather2019}. In addition, some of the most commonly used pre-trained networks have been shown to have redundancy in their feature representations when reconstructing for the best-perceived quality \cite{tariq2020}. The losses defined on those feature representations improve the perceived quality in the fovea but are not specifically optimized for peripheral vision. In this work, we take an orthogonal approach in the context of GAN training. Apart from using a perceptual loss to train a generator, our main contribution is a modification to the training of the discriminator such that it better reflects the discriminative power of a human observer.

\mysection{Perceptual experiments}{experiments_dataset}
There is an important connection between studies on metamers and those on foveated image reconstructions using GANs discussed above. While the former postulate the importance of preserving image statistics for peripheral vision, the latter reconstruct the content according to the discriminator trained on natural images and videos. However, we argue that training the discriminator using natural images does not adequately reflect the lack of sensitivity of the HVS to spatial distortions, and excessively constrain the generator in hallucinating content. Therefore, we propose to train the discriminator on a dataset composed of images that contain distortions that are unobjectionable to the observers.

It is important to note that the primary goal of this procedure is not to cause the GAN to produce distortions, but rather to make it insensitive to the distortions that humans cannot detect and focus on penalizing perceptually important artifacts. By doing this, we want the discriminator to share limitations similar to those of the HVS.

Training GANs on an extensive dataset of images containing distortions with a near-visibility-threshold requires the responses of human observers in a subjective experiment, in which the participants are asked to adjust the level of distortion in peripherally viewed stimuli. However, owing to the sheer amount of data typically required for training the GAN, it is unfeasible to generate such a dataset by relying solely on perceptual experiments in a controlled lab environment with a reasonable number of participants. Therefore, we rely on method of texture synthesis that takes advantage of the correlation between image features to preserve the statistical properties of an input image. By imposing pixel-level constraints, we can control how faithful the reconstruction is to the structure of an exemplar. The number of pixel-level constraints acts as a parameter that controls the freedom to change the structure of the synthesized image with respect to the input, thereby introducing a way to control the strength of visual distortions that can be permitted in the reconstruction. We aim to use subjective experiments to measure the strength of distortions, which makes the reconstruction metameric with respect to the peripherally viewed ground-truth stimulus. Once the optimal parameters have been estimated for a smaller set of images in the perceptual experiment, we generate a dataset large enough for training the GAN by using the method of texture synthesis, thus eliminating the need to conduct subjective experiments with an unreasonably large number of participants.

\mysubsection{Stimuli generation}{StimuliGeneration}
Our stimuli generation model is based on the texture synthesis method proposed by Gatys \etal \cite{gatys2015texture}.
We customized their method by partially constraining the synthesis to control the level of visual distortion in a convenient way. We exploited this capability to create metamers of the input image given the specific viewing conditions.
Their method was formulated as an optimization procedure on feature maps of the pre-trained VGG-19~\cite{vgg} network that optimizes $\hat{\vec{x}}$ for an input exemplar $\vec{x}$ by minimizing the loss function:

\begin{equation} \mathcal{L}(\vec{x}, \hat{\vec{x}}) = \sum_{l=0}^{L} w_l \sum_{i,j} \frac{1}{4N^2_l M^2_l}(\hat{G}^l_{ij} - G^l_{ij}) ,
\end{equation}

where $\hat{G}^l_{ij}$ and $G^l_{ij}$ are Gram matrices for feature maps $i$ and $j$ in layer $l$, $N_l$ is the number of feature maps, $M_l$ is the total number of neurons in a layer, and $w_l$ is an additional weight associated with layer $l$. To synthesize images for our experiment, we use the same procedure but also constrain a portion of randomly chosen pixel values in $\hat{\vec{x}}$ such that they are identical to the corresponding pixels in $\vec{x}$. We refer to these pixels as \emph{guiding samples}. We enforce the quality constraint by projecting the solution to the feasible space in each iteration of a gradient descent optimization.

\begin{wrapfigure}{R}{0.5\textwidth}
    \includegraphics[width=0.5\textwidth]{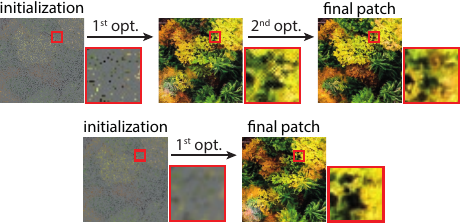}
    \caption{
    The figure presents two strategies for synthesizing stimuli by using the guiding samples. Top: a two-step procedure where a second unconstrained optimization is performed to remove subtle checkerboard artifacts after applying a Gaussian filter with $\sigma =1$. Bottom: constrained optimization initialized with a blurred version of the guiding samples.}
    \label{fig:texture_synthesis_suppl}
\end{wrapfigure}

\begin{figure}[b]
    \includegraphics[width=\textwidth]{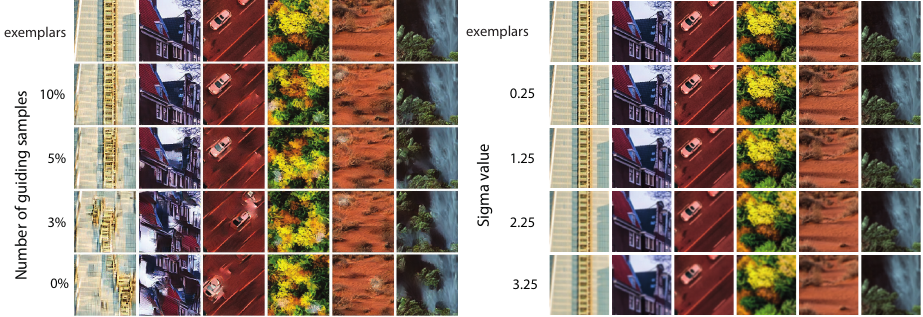}
    \caption{Results of stimulus generation for different numbers of guiding samples (left) and values of $\sigma$ (right) of the Gaussian filter. Note the increased distortions relative to the exemplar when the number of guiding samples decreases (left) and when $\sigma$ increases (right).}
    \label{fig:texture_synthesis}
\end{figure}

We observed that solving the constrained optimization leads to subtle image artifacts that resemble checkerboard patterns (\refFig{texture_synthesis_suppl}, first row -- middle). The problem is related to the well-known issue of checkerboard artifacts created by backpropagation \cite{Odena2016}. We identified two solutions to address this issue. The first consists of running the constrained optimization until convergence, removing the checkerboard artifacts characterized by a high spatial frequency by applying a low-pass Gaussian filter ($\sigma = 1$), and running a second round of optimization without the constraint. We also perceptually verified that similar results may be achieved when constrained optimization is initialized with the guiding samples filtered by a Gaussian filter (\refFig{texture_synthesis_suppl}, second row).

Using the above procedure, we computed the images $\hat{\vec{x_p}}$, where $p\%$ is the percentage of the guiding samples (see \refFig{texture_synthesis}, left). For $p=0$, our synthesis is equivalent to the original technique presented by Gatys \etal

The loss of high spatial frequencies (as commonly observed when reducing the input resolution) is an important factor influencing the perceived quality. To improve the sensitivity of the trained metrics to a visible decline in resolution, we also created a separate dataset of images with different degrees of Gaussian blur, i.e., different values of $\sigma$ of the Gaussian kernel (see \refFig{texture_synthesis}, right). The results of the perceptual experiment obtained with these stimuli were used to expand the dataset of images to train the image metrics in \refSec{image_quality_evaluation}.

\mysubsection{Experimental protocol}{perceptual_experiments}
The number of the guiding samples $p$ and the value of $\sigma$ of the Gaussian kernel provide a parametrization for our investigation of the sensitivity of the HVS to deviations relative to the original images. More precisely, with the generation of stimuli using texture synthesis in our main experiment, we sought a direct relationship between the number of guiding samples and the probability of detection of the distortions by a human observer at a particular eccentricity. We later used this relation to generate a much larger dataset of images with imperceptible distortions that is required for training GAN-based foveated reconstruction. By contrast, the additional experiment with Gaussian blur only sought pairs of images and the corresponding probability of blur detection, as a smaller dataset was sufficient for training our image metrics.

\paragraph{Stimuli}
We prepared 24 image patches of size $256 \times 256$\, each from a different 4K image. The images are grouped into two main categories: nature and architecture. Nature images typically do not contain as much structure as the architecture images of human-made objects. The features present in the nature images have a larger variance in their texture, both in terms of colors and frequency-related content. Natural objects have a large variety of shapes without any strict pattern. On the contrary, architecture scenes usually contain larger uniform areas with clear separation between different parts. We expected that reconstructing images with a clear structure may be more challenging.  For each patch, we generated a corresponding set of distorted patches for $p \in \{0, 3, 5, 7.5, 10\}$. The set of values of $p$ was determined in a preliminary experiment, in which we found that $p > 10\%$ yields to images that are almost always perceptually indistinguishable from the originals. A set of ground-truth patches was used to generate a set of blurred patches for $\sigma \in \{0.25, 1.25, 2.25, 3.25, 4.25\}$. The range was chosen to uniformly span the range of visible blur across the considered field of view \cite{Tursun2019}. The sample of our stimuli is presented in \refFig{texture_synthesis}.

\paragraph{Task}
The experiment started with a short initial warm-up phase, in which the participants received instructions about the task. Subsequently, in each trial, three patches were presented to them on the screen: (1) the original patch at the fixation point, (2) a synthesized stimulus on either the right or the left side at a given eccentricity, and (3) the original patch on the opposite side at the same eccentricity. The stimuli were visible to both eyes, and the participants were asked to select the patch that was more similar to the reference by pressing left or right arrow keys of the keyboard.

Although we asked the participants to maintain their gaze at the center of the screen during the experiment, involuntary changes in the position of the gaze might have occurred from time to time. To preserve the retinal position of the stimuli against involuntary movements of the eyes of the participants, the stimuli followed the eye movements (see \refFig{gazes}). For this purpose, the gaze position is continuously monitored by using an eye tracker. The participants were not required to always focus on one point because the stimuli always followed the gaze point. The participants did not receive feedback on the correctness of their responses.
We tested the visibility of distortions at  \SI{8}{\degree} (the end of the parafovea \cite{wandell1997foundations}), \SI{14}{\degree} (the center of the perifovea \cite{wandell1997foundations, strasburger2011peripheral}), and \SI{20}{\degree}, for which the stimuli spanned  \SI{3.21}{\degree}, \SI{3.08}{\degree}, and \SI{2.89}{\degree}, respectively.
The distance between the participant and the display was set to 70 cm during the experiment. At this viewing distance, each pixel spanned approximately 0.012 visual degrees. We did not impose a limit on the viewing time, and the average duration of each trial was 2 seconds. In total, each participant performed 1800 trials, leading to a total duration of almost 1 hour. The order of the images, eccentricities, and sides on which the test stimuli were shown were randomized. We used these experimental settings with both types of stimuli (created using texture synthesis and Gaussian blur).

\paragraph{Hardware}
We used a setup consisting of a \SI{27}{\inch} Acer Predator display operating at a resolution of \SIrangebys{3840}{2160}{\pixel} at \SI{120}{\hertz} and a peak luminance of \SI{170}{\candela\per\meter\squared}. We used a Tobii Pro Spectrum eye tracker at a sampling rate of \SI{600}{\hertz} to track the position of the gaze.

\paragraph{Participants}
In order to investigate the potential effects of the participants' background and knowledge of the field, we conducted this experiment\footnote{approved by Ethical Committee of Università della Svizzera italiana, decision CE.2020.3.} with two groups of participants. The first group consisted of five participants (four of them were the authors) who had extensive experience in computer graphics and full knowledge of the task. The second group consisted of 10 naive participants who had no experience in computer graphics or related fields. To improve the diversity of the stimuli, the naive participants performed the experiments with an extended dataset that contained four additional images. All participants were between 25 and 36 years of age and had a normal or corrected-to-normal vision.

\begin{figure}
    \includegraphics[width=\textwidth]{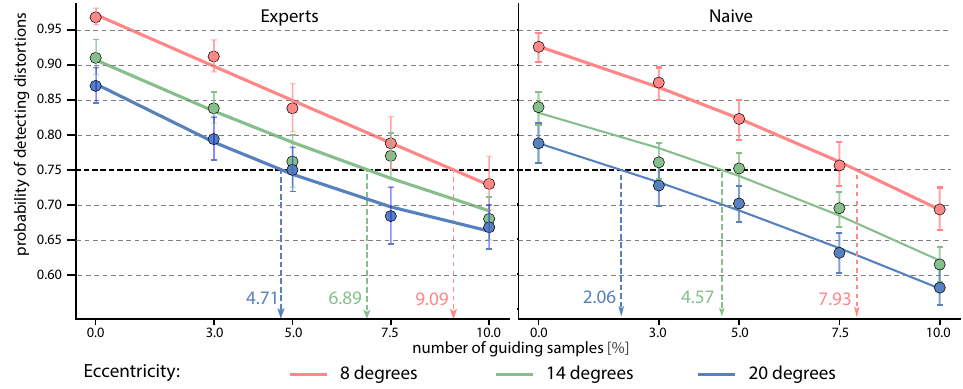}
    \caption{The probability of detecting differences between the original and the synthesized images as a function of the number of guiding samples and eccentricity. The error bars visualize standard errors of the mean. On the left are the results of the expert group and on the right are those of the naive group.}
    \label{fig:detection_sampling_rate_fig}
\end{figure}

\subsection{Data analysis}
The results of Gaussian blur were used directly to train the image metrics (\refSec{image_quality_evaluation}). The rest of the data, i.e., for texture synthesis-based stimuli, were processed separately for the expert and naive participants, and then compared. For each eccentricity and number of guiding samples tested, we first aggregated the participants' answers across all images and then computed the probability of their detection of distortions in the patches. We then expressed the relationship between the number of guiding samples and the probability of detection for each eccentricity by using a cubic polynomial fit. The results are shown in \refFig{detection_sampling_rate_fig} for both the expert and naive groups.

To construct a set of image patches to train a discriminator in a GAN architecture, we sought a relation between eccentricity and the number of guide samples, to produce patches that did not contain objectionable artifacts when used with texture synthesis. We used the probabilities obtained from the experiments for guidance. More precisely, we used the number of the guiding samples corresponding to a 75\% probability. Our choice was motivated by the commonly used definition of one just-noticeable-difference (JND) being a transition between visible and invisible distortions \cite{Mantiuk2021}. In practice, this is the midpoint between distortions that are always visible and those that are invisible. Although lower probabilities, such as 50\%, can be considered, we argue that 75\% provides a good trade-off for two reasons. First, the probabilities estimated from psychological experiments can asymptotically converge to only 50\%, posing challenges when seeking the exact 50\% point. The estimation of the threshold becomes ill-conditioned when the psychometric slope approaches zero, while 75\% is an adequately located value as it lies in the steepest part of the psychometric function. Second, our experiments applied an isolated scenario in which the participants were given the particular task of determining the higher-quality patch, and only the reference and the test patch were shown to them. In ultimate applications such as foveated rendering, the observer is less sensitive to any distortion. Therefore, we believe that choosing the number of guiding samples leading to near-threshold distortions was appropriate in this case.

We used a 75\% probability as a guide together with cubic fits to our data. We computed the number of guiding samples for expert subjects as 9.09\%, 6.89\%, and 4.71\% with a 95\% CI [7.85 - 10.48], [5.78 - 8.14], [3.60 - 5.94] for eccentricities \SIlist{8;14;20}{\degree}, respectively. For the naive participants, the results are: 7.93\%, 4.57\% and 2.06\% with a 95\% CI [7.47 - 8.41], [3.98 - 5.29], [1.30 - 2.76]. The confidence intervals were computed by bootstrapping, and the estimates are shown in \refFig{detection_sampling_rate_fig}. As expected, the naive observers were less sensitive to the artifacts than the experts, and tolerated distortions in synthesized patches with fewer guiding samples. We used the estimated values to prepare the inputs for the discriminator during GAN training (\refSec{dataset}).

\mysection{Method}{method}
The results of the perceptual experiments described in the previous section provide the measured thresholds for structural distortions for a standard observer. We used these data to control the learned manifold of the target images in foveated image reconstruction. To this end, we present an improved training scheme for the GAN in which the training data consist of a set of natural and synthesized images.

\begin{figure}
    \centering
    \includegraphics[width=\columnwidth]{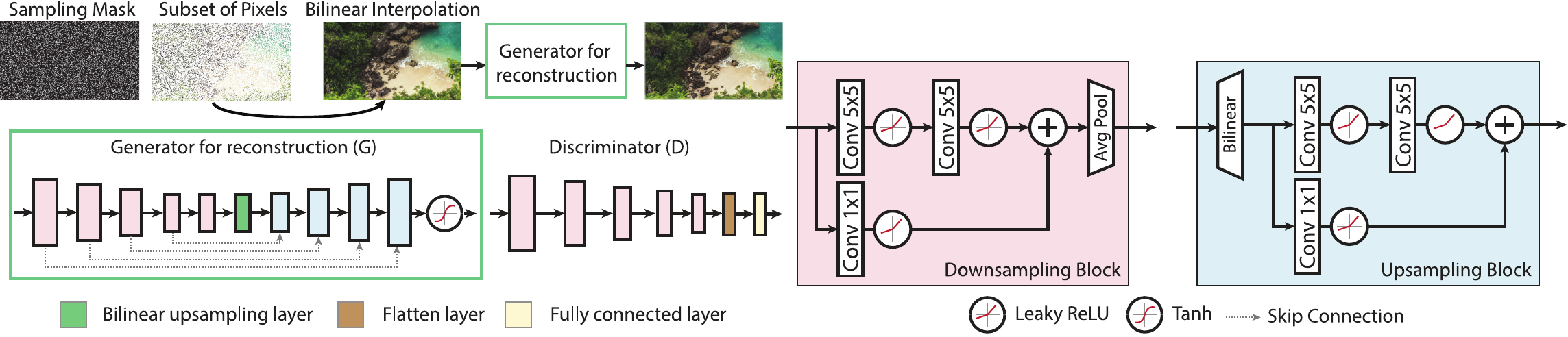}
    \caption{The architecture of our proposed network.}
    \label{fig:unet}
\end{figure}

Our network for the foveated reconstruction uses the Wasserstein GAN~\cite{wgan} training scheme to produce perceptually optimized reconstructions from subsampled images. The network topology is based on the UNet encoder-decoder structure with skip connections \cite{unet} (\refFig{unet}). This network design is similar to the model previously used by Kaplanyan \etal \cite{deepfovea}. The encoder part of the generator network ($G$) consists of downsampling residual blocks that use average pooling layers \cite{he2016}. Each residual block of the encoder consists of two convolutional layers with a filter size of $ 5 \times 5 $, except for the main branch, where we use a $ 1 \times 1 $ filter to adjust the dimensionality. The numbers of filters are 16-32-64-128-128 in each block, respectively. The decoder part is a mirrored version of the first four encoder blocks with upsampling blocks that use bilinear interpolation instead of average pooling. The encoder and the decoder are connected by an additional bilinear upsampling layer. We use LeakyReLU activation with a negative slope coefficient of $ \alpha = 0.2 $ throughout the network, except for the final layer of the generator, which uses $ \tanh $ activation.

The discriminator network or \emph{critic} ($D$) is based on PatchGAN \cite{patchgan}, with a patch size of $ 64 \times 64 $. The discriminator consists of downsampling blocks similar to the encoder part of the generator (the number of filters: 16-32-64-128-128). The output of the downsampling blocks is flattened and passed to a fully connected layer that produces a scalar. Compared with the model developed by Kaplanyan~\etal, we use a more compact generator with half the number of filters in the first and last three blocks. Such a compact generator is made possible by our more permissive training scheme, which allows for imperceptible deviations from the statistics of the target image. By contrast, the discriminator loss used by Kaplanyan~\etal aims to match the statistics of the target image as closely as possible. This important difference in our training scheme makes it possible to retain the perceptual quality of the image with a more compact network.
Furthermore, their technique aims to reconstruct the original image, while our work aims to show the feasibility of using a GAN trained on perceptual data.

\mysubsection{Dataset}{dataset}

We used two separate datasets as inputs to the generator and the discriminator. The input dataset for the generator consisted of patches from natural images with a size of $256\times256$. The patches were generated by cropping images with random offsets. To maintain a balanced data representation, 50 images were randomly selected from each of the 1000 classes in the ImageNet dataset \cite{deng2009imagenet}, which provided us with a total of 50K patches. These images were later sub-sampled using the void-and-cluster algorithm \cite{ulichney1993void} with sampling rates of $12\%$ for the near periphery and $0.7\%$ for the far periphery. This choice was made according to a content-aware foveated rendering method proposed by Tursun \etal \cite{Tursun2019}. The sub-sampling was followed by bilinear interpolation before the images were passed on to the generator as input.

We used the texture synthesis method described in \refSec{StimuliGeneration} using the results of the perceptual experiment described in \refSec{perceptual_experiments}. This dataset consisted of 50K patches that were synthesized using \SI{9.09}{\percent} and \SI{6.89}{\percent} of the pixels as guiding samples, respectively, for near and far peripheral regions in addition to the full-resolution ground truth images.

\mysubsection{Discriminator loss}{improved_adv_loss}

The training of the discriminator, $D$, uses the same loss function as in the original WGAN design \cite{wgan}:
\begin{equation}
	\mathcal{L}_{adv} = D(x) - D(G(z)),
\end{equation}
where $z$ represents the input to the generator network $G$, $D(x)$ is the output of the discriminator to real samples (natural images), and $D(G(z))$ is the output of the discriminator to reconstructions from $G$. This training is equivalent to the optimizations performed in previous work when $x \in \mathbb{I}$, where $\mathbb{I}$ is the set of images from ImageNet.

In our experiments, we updated this formulation by using $x^* \in \mathbb{I}^*$, where $\mathbb{I}^*$ is the set of images with visually imperceptible structural distortions, as we described in \refSec{perceptual_experiments}. We denote the discriminator loss operating on this manifold of synthesized images with structural distortions by:
\begin{equation}
 \mathcal{L}_{adv}^* = D(x^*) - D(G(z)).
\end{equation}
To ensure the stability of the training of the WGAN, we imposed a soft Lipschitz constraint by using a gradient penalty \cite{improved_wgan}.

\mysubsection{Generator loss}{generator_loss}
Our optimization trained generator networks with a weighted sum of different types of losses. For a comprehensive evaluation of our training scheme, we focused our analysis on three types of generators trained with standard and perceptual losses. The first generator, $G_{L2}$, was trained with a combination of standard MSE loss and adversarial loss.
The second generator, $G_{LPIPS}$, was trained with the learned perceptual image patch similarity loss term. We used the learned linear weights on top of the VGG network as provided by the authors in their work \cite{zhang2018unreasonable}.
Moreover, inspired by \cite{hepburn2019enforcing}, we added the generator $G_{Lapl}$ that used Laplacian-based loss. It is defined as a weighted sum of the mean squared error between the corresponding levels of the Laplacian pyramid for the reconstruction and the ground truth. We assigned weights to each level according to a Gaussian with $\sigma=1.0$. By centering the Gaussian around different levels, we were able to place more emphasis on the reconstruction fidelity of different spatial frequencies in the pyramid decomposition. The main motivation behind the loss was that by assigning a larger weight at lower spatial frequencies, the network will be given more freedom to hallucinate high spatial frequencies, which might be desirable in the periphery.

All the generator losses used in our experiments are listed in \refTbl{methods}.

\mysubsection{Training}{training_procedure}

\begin{figure*}
    \includegraphics[width=\textwidth]{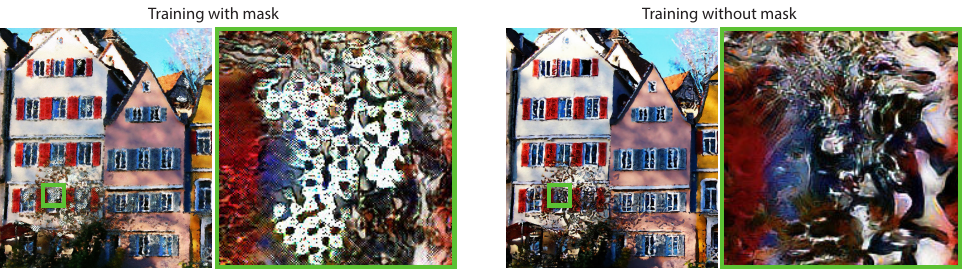}
    \caption{Training with (left) and without (right) an input sampling mask.}
    \label{fig:polka_dots}
\end{figure*}
Inspired by work by Guenter \etal \cite{guenter_foveated_2012}, we considered a foveated rendering scenario in which the image was divided into three regions with different levels of distortions. We assumed that the boundaries of these three regions were at the \SI{8}{\degree} and \SI{14}{\degree} of eccentricity (represented as the red and green circles, respectively, in \refFig{samples_visualization}), which coincide with the end of the parafovea and its center of perifovea \cite{wandell1997foundations}. They divided the image into foveal, near periphery, and far periphery regions. Our split was similar to that used in the computer graphics research, such as by [Guenter et al. 2012, Patney et al. 2016].

While the content of the foveal region was directly transferred from a full-resolution image to ensure the highest quality, we reconstructed the near and far periphery regions from a sparse set of samples. The input sampling density for these regions was assumed to be the same as the number of guiding samples estimated in our experiment with image patches (\refSec{experiments_dataset}). More precisely, we assumed that the near periphery region was reconstructed from the number of samples required for an eccentricity of \SI{8}{\degree}, and the far periphery was reconstructed from the number of samples corresponding to an eccentricity of \SI{14}{\degree}.

To perform this reconstruction, we trained two distinct generator networks, each of which was responsible for the reconstruction of the near and far peripheral regions. For benchmarking purposes, we also trained separate networks for each of the two discriminator losses (using our $\mathcal{L}_{adv}^*$ and the standard loss $\mathcal{L}_{adv}$) and three generator losses (using $\mathcal{L}_G^{L2}$, $\mathcal{L}_G^{LPIPS}$, and $\mathcal{L}_G^{Lapl}$). The relative weights of the loss terms were set to $ w_{L2} = 2000 $, $ w_{LPIPS} = 100 $, $ w_{Lapl} = 100 $, and $ w_{adv} = 1 $. The selected weights were adjusted according to the magnitudes of the individual loss terms to equalize their contributions to the final loss. This was done by observing the values of the individual terms during training.

We used the Adam optimizer with a learning rate of $2\times10^{-5}$ ($\beta_1 = 0.5 $, $\beta_2 = 0.999 $, $ \epsilon = 10^{-8} $). The training lasted for 20--30 epochs until convergence, which took approximately one day on an Nvidia 2080 Ti GPU. We assumed convergence when the training loss reached a plateau. The sample reconstructions from the converged network were also visually checked against potential instabilities during training.

\mysubsection{Sampling mask}{sampling_mask}
By capitalizing on the potential correlation between subsequent frames, the network introduced by Kaplanyan \etal uses recurrent connections as an important part of their design to retain information from previously subsampled frames. This high-level temporal reprojection provides the network with additional information when the underlying content is only partially observed owing to sparse subsampling. In order to clearly observe the effects of different training schemes, we used information from only one frame and isolated the reconstruction from the effects of this flow of temporal information. In our initial experiments, we observed that such a design decision made the network more sensitive to the sampling mask used in the inputs due to the absence of the flow of temporal information, which would otherwise have compensated for the lack of information on the true values of the missing samples. In order to address this issue, as a first attempt, we filled in the missing information by interpolating the sampled pixels while retaining the sampling mask as a channel of the input. However, visual inspection revealed visual artifacts collocated with the sampled pixels (\refFig{polka_dots}), and their visibility was dependent on the weights assigned to the loss terms. The effect was the most pronounced when we used $\mathcal{L}_{L2}$ in training, and the artifacts were less visible with $\mathcal{L}_{LPIPS}$.
As a remedy, we removed the sampling mask from the training input and provided the generator with the bilinearly interpolated input consisting of RGB channels, as shown in \refFig{unet}. This solution seemed to be effective in removing visual artifacts from the reconstruction (please refer to \refFig{polka_dots} for a visual comparison).

\mysection{Results and Evaluation}{experiments}
We evaluated our strategy for training foveated image reconstruction using objective image metrics (\refSec{image_quality_evaluation}) and a subjective experiment (\refSec{subjective_experiment}). In our evaluation, we aimed to show that the benefits of our method are not limited to a particular selection of the training loss. To this end, we evaluated the generator network $G$ of our method with six loss functions that were combinations of LPIPS ($\mathcal{L}_{LPIPS}$), L2 ($\mathcal{L}_{L2}$), and Laplacian pyramid loss ($\mathcal{L}_{Lapl}$) terms (\refTbl{methods}). For the discriminator network, $D$, we benchmarked the performance of networks trained using our new patch dataset ($ \mathcal{L}_{adv}^* $) as well as the original dataset ($ \mathcal{L}_{adv} $).

\begin{table}
	\centering
	\caption{The loss functions used for training the generator in our evaluations.}
	\label{tbl:methods}
	\begin{tabular}{ll}
		\toprule
		Loss function & Definition \\
		\midrule
		\rule[0ex]{0pt}{2.5ex} L2 & $\mathcal{L}_G^{L2} = w_{L2} \cdot \mathcal{L}_{L2} + w_{adv} \cdot \mathcal{L}_{adv}$ \\
		\rule[-1ex]{0pt}{2.5ex} L2 ours & $\mathcal{L}_{G^*}^{L2} = w_{L2} \cdot \mathcal{L}_{L2} + w_{adv} \cdot \mathcal{L}_{adv}^*$ \\
		\rule[-1ex]{0pt}{2.5ex} LPIPS & $\mathcal{L}_G^{LPIPS} = w_{LPIPS} \cdot \mathcal{L}_{LPIPS} + w_{adv} \cdot \mathcal{L}_{adv}$ \\
		\rule[-1ex]{0pt}{2.5ex} LPIPS ours & $\mathcal{L}_{G^*}^{LPIPS} = w_{LPIPS} \cdot \mathcal{L}_{LPIPS} + w_{adv} \cdot \mathcal{L}_{adv}^*$ \\
		\rule[-1ex]{0pt}{2.5ex} Laplacian & $\mathcal{L}_G^{Lapl} = w_{Lapl} \cdot \mathcal{L}_{Lapl} + w_{adv} \cdot \mathcal{L}_{adv}$ \\
		\rule[-1ex]{0pt}{2.5ex} Laplacian ours & $\mathcal{L}_{G^*}^{Lapl} = w_{Lapl} \cdot \mathcal{L}_{Lapl} + w_{adv} \cdot \mathcal{L}_{adv}^*$ \\
		\bottomrule
	\end{tabular}
\end{table}

\mysubsection{Visual inspection}{visual_inspection}

\begin{figure*}
    \centering
    \includegraphics[width=\textwidth]{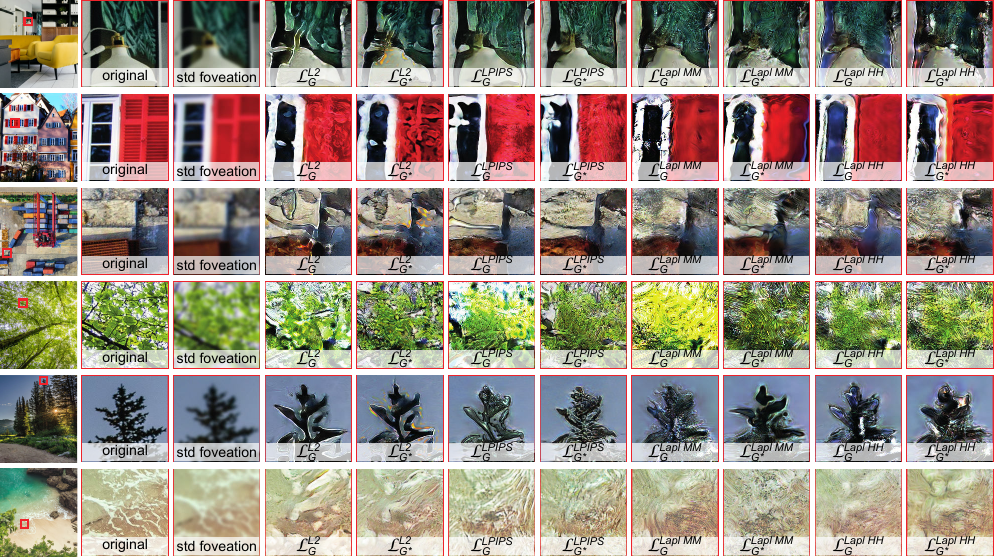}
    \caption{Sample reconstructions of nature and architecture images by all evaluated methods for the far periphery. \textit{$\mathcal{L}_G^{Lapl\,MM}$} refers to the largest weight assigned to medium spatial frequencies, while \textit{$\mathcal{L}_G^{Lapl\,HH}$} refers to the assignment of the largest weight to high spatial frequencies. Note that the provided images are shown for the demonstration purposes only. For appropriate perception cues, the images corresponding to the figure need to be observed in full screen with 36 cpd. They correspond only to the far peripheral area of vision, spanning from 21° to the corners of the display.}
    \label{fig:discriminator_input}
\end{figure*}

\refFig{discriminator_input} presents the results of reconstruction obtained by using differently trained architectures on four images. For reference, we include the original high-resolution and the standard foveated reconstruction by using an interpolation with Gaussian weights. For the results of training using Laplacian loss, we introduced a notation consisting of two letters, $\mathcal{L}_G^{Lapl\,XY}$, where X, Y $ \in \lbrace \text{H}, \text{M} \rbrace $ encode the position of the Gaussian peak at the far and near periphery, respectively. The letter H represents the position of the peak located at the first level of the pyramid (characterized by an emphasis on high spatial frequencies), whereas the letter M represents the position of the peak located at the fourth level of the pyramid (medium spatial frequencies). For example, the method denoted by $\mathcal{L}_G^{Lapl\,HM}$ refers to a reconstruction obatined by using a network trained with Laplacian pyramid-based loss. In this case high spatial frequencies were assigned larger weights for the far periphery, and medium frequencies were given higher importance for the near periphery.

The first observation is that the results of all eight GAN-based reconstruction exhibit clear hallucinated results, and the reconstruction of very fine details is not exact. Although this is visible with direct visual inspection, such deviations are less visible when shown in the periphery. Furthermore, all reconstructions introduce high spatial frequencies and strong edges, but training with $\mathcal{L}_G^{L2}$ loss makes them sparser and more exaggerated. A visual comparison (\refFig{discriminator_input}) between the discriminators trained with and without our synthesized dataset (i.e., $\mathcal{L}_G^{L2}$ vs. $\mathcal{L}_{G^*}^{L2}$, $\mathcal{L}_G^{LPIPS}$ vs. $\mathcal{L}_{G^*}^{LPIPS}$, $\mathcal{L}_G^{Lapl\,MM}$ vs. $\mathcal{L}_{G^*}^{Lapl\,MM}$, $\mathcal{L}_G^{Lapl\,HH}$ vs. $\mathcal{L}_{G^*}^{Lapl\,HH}$) shows that our results include higher spatial frequencies. We argue that this is due to the flexibility of the discriminator, which penalizes hallucinations of high spatial frequencies less harshly. This is the desired effect because while the HVS is sensitive to the removal of some high spatial frequencies in the periphery, it is less sensitive to changes in their positions (\refSec{related_work}). %

\begin{figure*}
    \includegraphics[width=\textwidth]{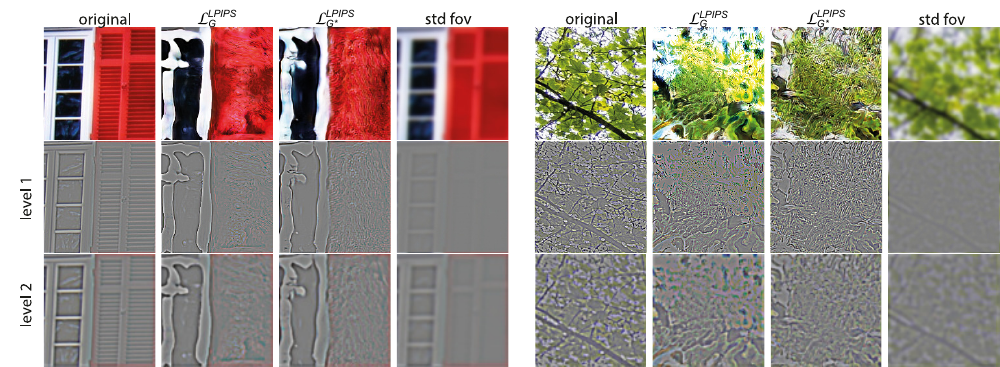}
    \caption{High-frequency content hallucination using $\mathcal{L}_G^{LPIPS}$, $\mathcal{L}_{G^*}^{LPIPS}$, and standard Gaussian blur.}
    \label{fig:differences_stdfov}
\end{figure*}

To further investigate the characteristics of spatial frequency distribution of our reconstructions, we visualized the outputs of frequency band decomposition of the Laplacian pyramid and computed the differences of two layers from the bottom of the pyramid\footnote{Please note the difference between frequency decomposition using the Laplacian pyramid and the Laplacian pyramid-based loss function.}, which encode the highest frequency band as well as one octave below it (\refFig{differences_stdfov}). We observe that our reconstructions provide additional hallucinated high-frequency details that do not exist in the traditional foveated image reconstruction.
Please refer to the supplementary material for an interactive demo with more results.

\mysubsection{Objective image metrics}{image_quality_evaluation}

We assessed the perceptual quality of our foveated image reconstruction using the recently introduced FovVideoVDP metric \cite{Mantiuk2021}. FovVideoVDP is a full-reference quality metric that can be used on images and videos. It considers into account the peripheral acuity of the HVS and the retinal eccentricity of the stimuli while computing quality scores. FovVideoVDP quality scores are in Just-Objectionable-Difference (JOD) units ($\text{JOD} \in [0, 10]$) where $ \text{JOD} = 10 $ represents the highest quality, while lower values represent higher perceived distortion with respect to the reference. We computed FovVideoVDP quality scores of the images generated by our method ($ \mathcal{L}_{adv}^* $) and those reconstructed by networks trained on a standard dataset ($ \mathcal{L}_{adv} $). We provided the original image to the metric as a reference image. We report the FovVideoVDP quality scores in \refTbl{fovvdp} for different peripheral regions (near and far), and generator loss functions. We compared these scores with those of our training method using $ \mathcal{L}_{adv}^* $ and the standard training approach with $ \mathcal{L}_{adv} $. Our method achieved higher quality-related scores than the standard approach to training the GAN. The generator was able to reconstruct the images better when we included perceptually non-objectionable distortions in the training set of the discriminator using our method.

\begin{table}[h!]
\centering
\caption{FovVideoVDP \cite{Mantiuk2021} quality scores (in JOD units) of our method and standard training of GANs for image reconstruction. The scores were computed for near and far peripheral regions, which represent the images reconstructed at \SI{8}{\degree} and \SI{14}{\degree}, respectively. A higher score implies better visual quality.}
\label{tbl:fovvdp}
\begin{tabular}{cccc}
\toprule
Peripheral region & Loss function & Ours & Standard \\
\midrule
\multirow{2}{*}{Near (\SI{8}{\degree})} & $\mathcal{L}_G^{L2}$    & \textbf{8.25} & 8.08     \\[5pt]
                      & $\mathcal{L}_G^{LPIPS}$ & \textbf{8.19} & 8.11     \\
\midrule
\multirow{2}{*}{Far (\SI{14}{\degree})}  & $\mathcal{L}_G^{L2}$    & \textbf{6.44} & 6.28     \\[5pt]
                      & $\mathcal{L}_G^{LPIPS}$ & \textbf{6.31} & 6.18     \\
\bottomrule
\end{tabular}
\end{table}

We also evaluated our method by using other objective quality metrics. Although many objective quality metrics are available for non-foveated quality measurement, objective quality assessment for foveated images is still an open research problem. In the absence of quality metrics for specific types of image distortions, past work has shown that the task-specific calibration of currently available objective quality metrics may be a promising solution \cite{wolski2019, adhikarla2017}. Motivated by this, we used our perceptual data to calibrate existing metrics: L2, SSIM \cite{ssim}, MS-SSIM \cite{msssim}, and LPIPS \cite{zhang2018unreasonable}, separately for different eccentricities. The calibration was performed by fitting the following logistic function \cite{richards1959flexible}:
\begin{equation}
\label{logistic_model}
    y(t) = a + (k - a)/(c + q \cdot e^{-b\cdot t})^\frac{1}{v}.
\end{equation}

to reflect the non-linear relation between the magnitude of distortion in the image and the probability of detecting it, with $a,b,c,k,q,v$ being free parameters. Inspired by LPIPS \cite{zhang2018unreasonable}, we also considered reweighing the contributions of each convolution and pooling layer of VGG-19 for each eccentricity separately. We refer to this metric based on the calibrated VGG network as Cal.\,VGG.

For all metrics, the free parameters (i.e., the parameters of the logistic functions as well as the weights and bias of VGG-19 layers) were obtained by minimizing the mean squared error in predicting the probability of detection:
\begin{equation}
    \sum_{(\vec{x},\hat{\vec{x}})\in S_r} \left\lVert y(M(\vec{x},\hat{\vec{x}})) - P(\vec{x},\hat{\vec{x}})\right\lVert^2,
\end{equation}
where $M$ is one of the original metrics, $S_r$ is the set of distorted and undistorted pairs of images for eccentricity $r \in {8,14,20}$, and $P$ is the probability of detecting the difference. Minimization was performed by using nonlinear curve fitting through the \textit{trust-region-reflective} and the \textit{Levenberg-Marquardt} optimizations \cite{branch1999subspace,levenberg1944method} with multiple random initializations. Furthermore, we constrained the VGG weights to be non-negative values to maintain the positive correlation between image dissimilarity and the magnitude of differences in VGG features, as motivated by the work in~\cite{zhang2018unreasonable}. %
To make our dataset more comprehensive, we added stimuli from an additional experiment that analyzed the visibility of the blur. For this purpose, we followed the procedure described in \refSec{perceptual_experiments}. %

\begin{wrapfigure}{R}{0.5\textwidth}
    \includegraphics[width=0.5\textwidth]{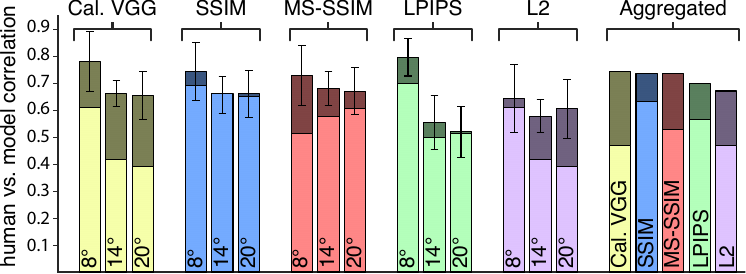}
    \caption{Pearson's correlation coefficient of the analyzed methods. Bright bars: uncalibrated metrics (using standard logistic sigmoid with the equation: $y(t) = 1/(1+e^{-t})$), dark bars: calibrated metrics (fitted to the logistic model using Eq.~\ref{logistic_model}).}
    \label{fig:quality_methods_compared}
\end{wrapfigure}

To validate our calibration, we performed 5-fold cross-validation and computed Pearson's correlations between the ground-truth probability of detecting distortions and metric predictions. \refFig{quality_methods_compared} presents correlation coefficients for all trained metrics and eccentricities computed as an average across all the folds. Each bar shows the measured correlation for the uncalibrated (bright part) and calibrated (dark part) metrics by using the data from our initial experiment (\refSec{perceptual_experiments}). For uncalibrated metrics, we used the standard sigmoid logistic function: $y(t) = 1/(1+e^{-t})$. We also provide the aggregated results, where the correlation was analyzed across all eccentricities.
The individual scores showed that as the eccentricity increase, there is a decline in performance in terms of the original metrics. The additional calibration significantly improves the prediction performance in terms of all metrics.
An interesting observation is that LPIPS performs very well for small eccentricities (\SI{8}{\degree}). For larger ones (\SIlist{14;20}{\degree}), however, its performance is significantly reduced even with the optimized logistic function. We relate this observation to the fact that LPIPS is not trained for peripheral vision. However, when the weights of the deep layers of VGG-19 are optimized (Cal.\,VGG), the performance improved significantly. This suggests that the above metrics are promising, but depending on the eccentricity, the contributions of the individual layers to the overall prediction must change. Since our Cal.\,VGG delivered the best performance in the tests, we selected it to benchmark the foveated image reconstruction techniques listed in \refTbl{methods}. The results of this test for other metrics that we did not use for evaluation are also reported in the supplementary material as a reference.

\begin{wrapfigure}{R}{0.5\textwidth}
    \includegraphics[width=0.5\textwidth]{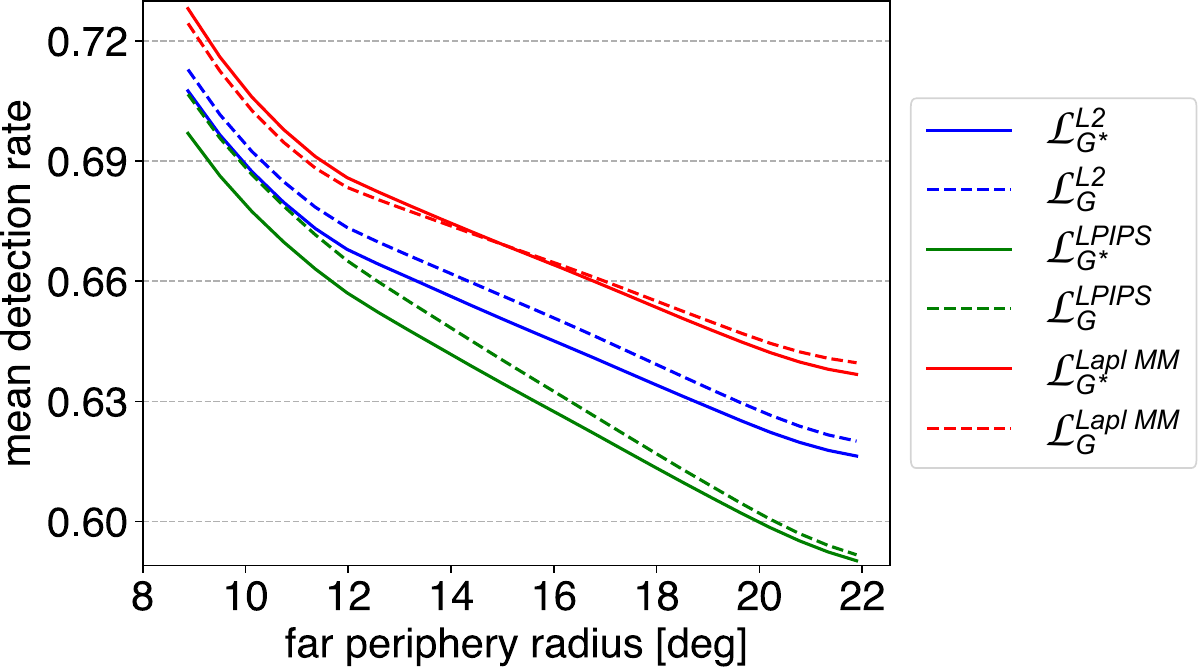}
    \caption{Detection rates according to the objective Cal.\,VGG metric with increasing radius of the far periphery. Lower values indicate a higher quality of reconstruction.}
    \label{fig:prediction_plot_vgg}
\end{wrapfigure}

After calibration, Cal.\,VGG is still limited to processing image patches as input. To be able to run Cal.\,VGG on full images that cover a larger field of view, it needs to consider the influence of changes in eccentricity depending on the position of a given pixel in the image. To support arbitrary values of eccentricity as input, we linearly interpolated the prediction of the metric from 8 and 20 degrees to intermediate values of the eccentricity between 8-20 degrees. Moreover, in contrast to our approach in the calibration step, we switched to a single logistic function whose parameters were estimated by using experimental data from all eccentricities. After these extensions, we ran Cal.\,VGG locally on non-overlapping patches of the full input image. To compute a single scalar for the entire image, we took the average value computed across all patches as a global pooling step. To benchmark different reconstruction methods, we randomly selected 10 publicly available images at $3840 \times 2160$ resolution that contained architectural and natural features. Before applying different reconstruction techniques, we split the images into three regions: fovea, near periphery, and far periphery. We then draw sparse samples as visualized in (\refFig{samples_visualization}). To test the reconstruction quality provided by different sampling rates used in near- and far-peripheral regions, we analyzed the Cal.\,VGG predictions for various blending strategies by changing the eccentricity thresholds at which the transition from near to far peripheral regions occurred. We computed the predicted detection rates from Cal.\,VGG for threshold points between \SI{9}{\degree}-\SI{22}{\degree} for all images. \refFig{prediction_plot_vgg} presents the results. Lower detection rates indicate lower probabilities of detecting reconstruction artifacts by human observers and, therefore, a higher reconstruction quality.
Training the reconstruction using $\mathcal{L}_{G^*}^{LPIPS}$ and $\mathcal{L}_G^{LPIPS}$ yields reconstructions that are the least likely to be distinguished from the original images. The results generated by using $\mathcal{L}_{G^*}^{L2}$ delivered a lower detection rate than those generated by using $\mathcal{L}_{G}^{L2}$. The detection rate for our method is significantly lower when the far periphery threshold is selected in the range of eccentricity of \SI{12}{\degree}-\SI{22}{\degree} ($p < 0.05$). For $\mathcal{L}_{G*}^{LPIPS}$, this difference in detection rate is significant, compared with that for $\mathcal{L}_{G}^{LPIPS}$ for thresholds between \SI{9}{\degree}-\SI{16}{\degree} ($p < 0.05$). We did not note a significant difference between the methods ($p > 0.10$ for all cases considered) when the network was trained by using Laplacian loss. All $p$-values were computed by using $t$-test.

We separated the images into two groups according to their prominent visual features: nature and architecture. Nature images were considered to form a class containing fewer geometrical structures and more texture-like areas, e.g., leaves, trees, waves, etc. They usually have a large variety of shapes without any well-defined patterns. In addition, they exhibit a high level of variance in colors and structure. Visual distortions in nature images would be less likely to result in perceivable changes because the variance in color and structure may have a masking effect on the distortions. %
On the contrary, architecture images mostly contain structures, like human-made objects, such as buildings, and larger uniform areas with clear visual boundaries. They usually have many edges and corners, which makes it more challenging to have a perceptually plausible and faithful reconstruction from sparse image samples. Distorting such images is more likely to lead to the mixing of visual information from different areas, and this is easy to detect even in the peripheral region of vision. Owing to these distinct properties of nature and architecture images, we separately evaluated the results on these two types of images separately. The results show that the difference of detection rate between our method and the standard training, compared to the overall trend, is more pronounced for nature images and less pronounced for architecture images. The results are available in Figure 6 of the supplementary material.

\mysubsection{Subjective experiments}{subjective_experiment}
The psychovisual experiment used to derive the data for training our reconstruction methods was performed involved 5 participants. Although it is common to use few participants in such experiments due to their complexity, and given that they should capture the general properties of the HVS, such experiments do not investigate potential differences within the population. In addition, it is not clear whether the method derived from the perceptual data is effective. Therefore, to further validate our claims regarding the new training strategy, and verify the importance of the improvements observed when using calibrated metrics, we conducted an additional subjective user experiment, in which naive participants were asked to directly compare different reconstruction methods.

\begin{figure*}[b]
    \includegraphics[width=\textwidth]{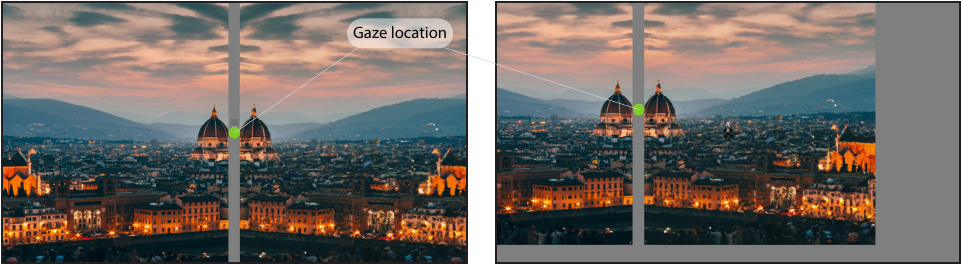}
    \caption{An example of the stimuli shown during the experiment. The image follows the gaze point, which is marked with the green dot.}
    \label{fig:gazes}
\end{figure*}

\paragraph{Stimuli and task}
We used the 10 images that were used in our evaluation on objective image metrics (\refSec{image_quality_evaluation}). They were sub-sampled and reconstructed by using $\mathcal{L}_{G^*}^{L2}$, $\mathcal{L}_{G}^{L2}$, $\mathcal{L}_{G^*}^{LPIPS}$, and $\mathcal{L}_G^{LPIPS}$ as shown in \refFig{samples_visualization}. In each trial, the participants were shown the original image on the left and one of two reconstructions on the right half of the display. The two halves were separated by a 96-pixel-wide gray stripe. The participants could freely switch between reconstructions by using a keyboard. They were asked to select the reconstruction that was more similar to the reference on the left by pressing a key. During the experiment, the images followed the eye movements of the participant, as shown in \refFig{gazes}. In contrast to the calibration experiment performed in \refSec{perceptual_experiments}, in this experiment we showed full images to the participants, each covering half of the screen. Fixation was enforced as in the calibration experiment described in \refSec{perceptual_experiments}. Each trial took 15 seconds on average. The total duration of the experiment was around 15 minutes.

\paragraph{Participants}
15 participants with normal or corrected-to-normal vision took part in the experiment. All were naive for the purposes of the study and were given instructions at the beginning. Each participant was asked to compare all pairs of techniques for each image (60 comparisons per participant).

\paragraph{Results}
To analyze the results, for each pair of techniques, we computed a preference rate of method A over method B. The rate expresses the percentage of trials in which method A was chosen as visually more similar to the original image. \refTbl{preferences_ours} shows the preference rates obtained by using networks trained using our procedure ($\mathcal{L}_{G^*}^{LPIPS}$, $\mathcal{L}_{G^*}^{L2}$) in comparison with the standard procedure ($\mathcal{L}_G^{LPIPS}$, $\mathcal{L}_G^{L2}$). We report the results for all images (last column) and do so separately for nature and architecture images. We used the binomial test to compute the $p$-values. The reconstructions obtained by using a network trained with our strategy are preferred in 57\% of the cases ($p = 0.013$). The difference is significant for $\mathcal{L}_{G^*}^{LPIPS}$ with a 59\% preference ($p = 0.04$). In the context of different image classes, our method performed well on nature images, both when we consider $\mathcal{L}_{G^*}^{L2}$ and $\mathcal{L}_{G^*}^{LPIPS}$ separately and when we consider them jointly (for each case preferred in 75\,\% of cases, $p < 0.001$). On architecture images, we observe the preference for $\mathcal{L}_G^{L2}$ (63\%, $p = 0.037$). For all techniques, our method is preferred in 40\,\% of the cases ($p = 0.018$). This is consistent with the results of Cal.\,VGG (\refSec{image_quality_evaluation}), where our method had a lower probability of detection of nature images and a similar probability of detection of architecture images. We hypothesize that there might be several reasons for its poorer performance on architecture images. First, architecture images contain objects with simple shapes, uniform areas, edges, and corners. Such features might not have been represented well in our calibration, where we used $256 \times 256$ patches, whose size was limited to avoid testing the visibility across a wide range of eccentricities. Furthermore, we believe that distortions in the visual features of simple objects are much easier to perceive than those in natural textures, which are more random. This problem might have been aggravated because our calibration considered both groups together and did not make any distinction when modeling the perception of artifacts for them. This problem might be solved by using different numbers of guiding samples for different classes of images when generating the dataset for training GAN-based reconstruction. However, this would require more careful data collection for the initial experiment and a more complex model that can predict the number of guiding samples based on the image content. Once these challenges have been addressed, the proposed approach can yield a more accurate dataset and can be used to train a single architecture that can handle different types of images.

\begin{table}
\centering
\caption{Preference rates of the methods $\mathcal{L}_{G^*}^{L2}$, $\mathcal{L}_{G^*}^{LPIPS}$ over $\mathcal{L}_G^{L2}$, $\mathcal{L}_G^{LPIPS}$ when trained using the data collected from the expert group. The values were computed by taking the average across participants. The errors correspond to the standard error of the mean.}
\label{tbl:preferences_ours}
    \begin{tabular}{ cccc }
    \toprule
    & Nature & Architecture & All \\
    \midrule
    $\mathcal{L}_{G^*}^{L2}$ vs $\mathcal{L}_G^{L2}$ & 0.75 $\mp$ 0.05 & 0.37 $\mp$ 0.04 & 0.56 $\mp$ 0.03 \\
    $\mathcal{L}_{G^*}^{LPIPS}$ vs $\mathcal{L}_G^{LPIPS}$ & 0.75 $\mp$ 0.05 & 0.43 $\mp$ 0.07 & 0.59 $\mp$ 0.05 \\
    All ours vs All & 0.75 $\mp$ 0.04 & 0.40 $\mp$ 0.04 & 0.57 $\mp$ 0.03\\
    \bottomrule
    \end{tabular}
\end{table}

\begin{figure*}
    \centering
    \includegraphics[width=\textwidth]{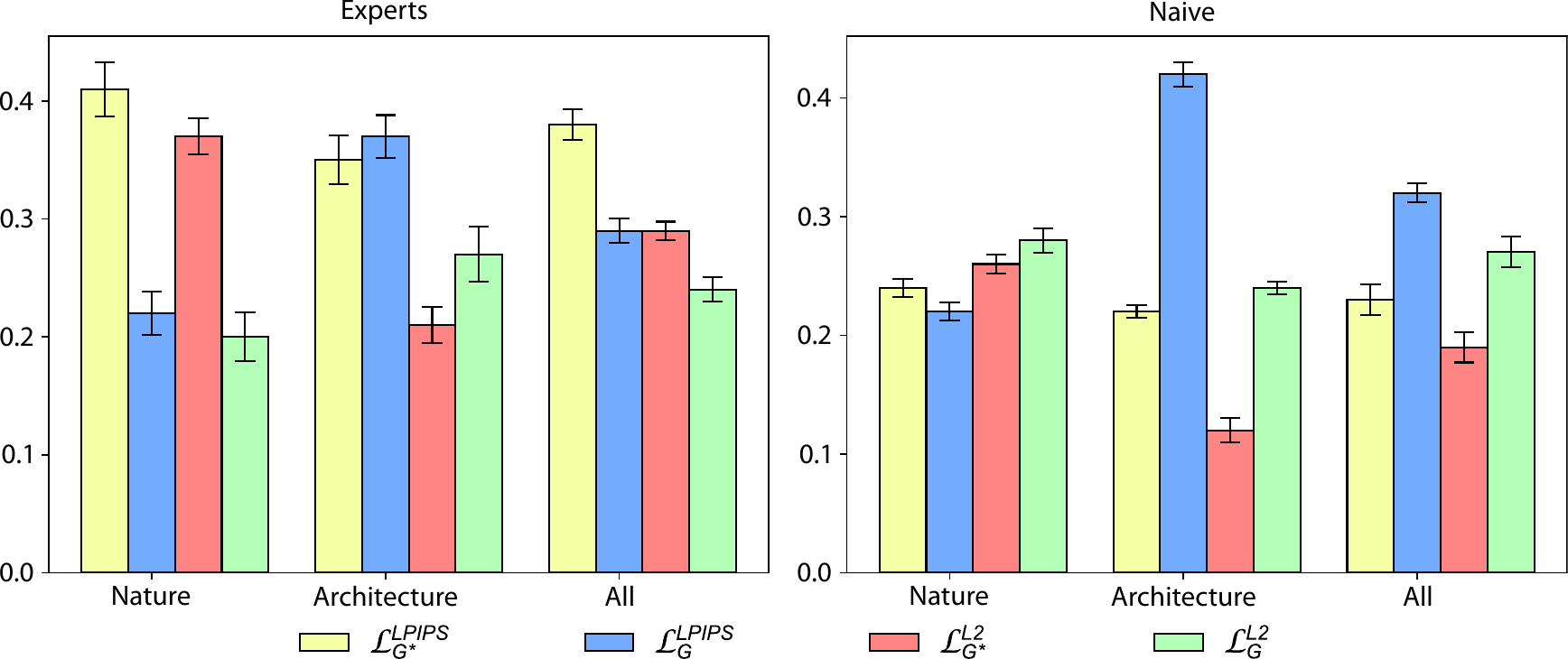}
    \caption{Preference rates of the methods trained using data gathered from the expert group (left) vs. the naive group (right) in the calibration experiment (\refSec{perceptual_experiments}). The scores on the $y$-axis represent the number of times a given method was selected, divided by the total number of trials of the experiment. The error bars show the standard error of the mean. The values were computed by taking an average across participants.}
    \label{fig:ranking_losses}
\end{figure*}

\refFig{ranking_losses} shows the preferences for the individual methods compared with those for all other training strategies, including different loss functions, i.e., $\mathcal{L}_{G^*}^{LPIPS}$, $\mathcal{L}_{G^*}^{LPIPS}$, $\mathcal{L}_{G}^{L2}$, and $\mathcal{L}_{G^*}^{L2}$. In the experiment with experts (left), $\mathcal{L}_{G^*}^{LPIPS}$ attained the highest preference of 38\% ($p < 0.001$) while $\mathcal{L}_G^{L2}$ recorded the lowest preference (24\%, $p < 0.001$). When divided into classes, $\mathcal{L}_{G^*}^{LPIPS}$ and $\mathcal{L}_{G^*}^{L2}$ were the most preferred methods for natural images, with values of 41\% ($p < 0.001$) and 37\% ($p = 0.003$), respectively. The other methods had lower preference values - 22\% for $\mathcal{L}_G^{LPIPS}$ ($p < 0.001$) and 20\% for $\mathcal{L}_G^{L2}$ ($p  < 0.001$). $\mathcal{L}_G^{LPIPS}$ was the most preferred on architecture images (37\%, $p = 0.005$), followed by $\mathcal{L}_{G^*}^{LPIPS}$ (35\%, $p = 0.032$). The $\mathcal{L}_{G^*}^{L2}$ was selected the fewest number of times (21\%, $p < 0.001$). All $p$-values were computed by using the binomial test, and the remaining results were not statistically significant.
The experiment, when repeated with naive participants (right), yielded a different threshold as a function of the guiding samples needed for the appropriate foveated reconstruction. In particular, the values related to \SI{8}{\degree} and \SI{14}{\degree} changed from 9.09 to 7.93, and from 6.89 to 4.57, respectively. This means that for a standard observer, the number of samples needed to generate an image of fixed quality is higher than the samples needed for an expert observer. Texture synthesis was the initial step of our pipeline. For this reason, we trained all our networks again and we repeated the validation experiment with the new reconstruction. The results are presented in \refTbl{preferences_ours_new} and \refFig{ranking_losses} (right). The new experiments showed that while our technique maintained a slight advantage through $\mathcal{L}_{G^*}^{LPIPS}$ over the standard method on nature images, our reconstructions delivered the worst performance on architecture images.

\begin{table}
\caption{Preference rates of the methods $\mathcal{L}_{G^*}^{L2}$, $\mathcal{L}_{G^*}^{LPIPS}$ over $\mathcal{L}_G^{L2}$, $\mathcal{L}_G^{LPIPS}$ when trained by using the data gathered from the naive group. The values were computed by taking the average across participants. The errors correspond to the standard error of the mean.}
\label{tbl:preferences_ours_new}
\begin{center}
\begin{tabular}{ cccc }
\toprule
& Nature & Architecture & All \\
\midrule
$\mathcal{L}_{G^*}^{L2}$ vs $\mathcal{L}_G^{L2}$ & 0.38 $\mp$ 0.05 & 0.32 $\mp$ 0.05 & 0.35 $\mp$ 0.02 \\
$\mathcal{L}_{G^*}^{LPIPS}$ vs $\mathcal{L}_G^{LPIPS}$ & 0.59 $\mp$ 0.04 & 0.17 $\mp$ 0.06 & 0.38 $\mp$ 0.04 \\
All ours vs All & 0.49 $\mp$ 0.03 & 0.24 $\mp$ 0.04 & 0.36 $\mp$ 0.03\\
\bottomrule
\end{tabular}
\end{center}

\end{table}

\subsection{Discussion}
The results of the final evaluation demonstrate that with a successfully derived training dataset containing near-threshold distortions, our GAN-based training strategy can improve the quality of the reconstructions. The benefits, however, are observed with a more conservative calibration (\refSec{experiments_dataset}) performed by experts. When calibration data from the naive participants are used, the final preference shifts towards the standard reconstruction strategy. This shows that our strategy works under conservative calibration conditions. We believe that this owes itself to the potential limitations of our calibration, which was performed on small patches, while larger patches may render some artifacts more prominent. However, there is a trade-off between using small patches and obtaining localized information about the sensitivity to distortions for a particular eccentricity, and making the patches larger and losing this property. In future work, it would be interesting to investigate better calibration strategies for our technique. 

As described, our technique can be directly applied to train GAN-based foveated image reconstructions. Our experiments demonstrate that the same network architecture with the same input provides higher-visual-quality reconstructions if our training strategy is used than otherwise. This applies directly to techniques such as the one proposed by Kaplanyan et al. \cite{deepfovea}. Because the density of the input sampling can be changed, we argue that our training can also reduce the number of input samples while preserving quality, which will improve the efficiency of the entire image generation pipeline.
\mysection{Conclusions and future work}{limitations}
Currently available techniques for foveated image reconstruction use perceptual loss to guide network training to capitalize on perceptually important image features. The goal of this work was to inject perceptual information into the discriminator network. To this end, during training, we provided the discriminator with images containing distortions that are imperceptible to a human observer. This allowed the discriminator to inherit the properties of the HVS encoded in the training dataset. Our new dataset contains images with invisible spatial distortions based on texture synthesis. We argue that such distortions are much closer to artifacts introduced by GAN-based reconstruction than the previously considered blur. Moreover, the new dataset allowed us to train several image metrics to improve their predictions of stimulus quality presented in the periphery.

We studied the suitability of the new training strategy for foveated image reconstruction. In future work, it is essential to extend this investigation to video content as this may yield benefits when the sensitivity of the HVS to temporal artifacts is incorporated. We trained separate networks for the near and far periphery. While this makes the training procedure easier, a more practical solution is to train one network to handle spatially varying density. An alternative solution to attain this goal is to use a fully convolutional network in the log-polar domain \cite{solari2012,wang2017}. We also did not focus on computational performance. At present, our unoptimized inference takes 3 seconds on our hardware. Although previous work \cite{deepfovea} has shown the feasibility of using GAN in such scenarios, computational efficiency remains an important concern. We believe that making networks and their training aware of the limitations of human perception will be important to close the gaps. 

Another exciting direction of research is to design a foveated image metric that accounts for a wide range of effects. While work by Mantiuk et al. \shortcite{Mantiuk2021} has taken this approach, they targeted image quality instead of the visibility of distortion. The challenge here lies in collecting large-scale perceptual data with eye-tracking-based information that can help determine the visibility of distortions. Even though our dataset contained this information, it is not sufficient to train a general-purpose visibility metric for both foveal and peripheral vision.

Finally, we believe that the idea of supplying the discriminator in the GAN architectures with images containing near-threshold distortions during training extends beyond applications to foveated image reconstructions. We see it as a more general strategy for training perception-aware GAN-based techniques for the creation of graphical content. Our work considered texture synthesis as a technique suitable for creating controlled distortions relevant to our application. However, it is not ideal for capturing other characteristics of perception, such as the sensitivity to temporal changes, color, and depth, that might be relevant in different applications. It would be interesting if other researchers followed our procedure and included these aspects in the training of generative-adversarial networks to verify their benefits.

\begin{acks}
This project has received funding from the European Research Council (ERC) under the European Union's Horizon 2020 research and innovation program (grant agreement N◦ 804226 PERDY).
The images used in this paper come from ImageNet dataset and Pexel.com. We would like to thank all who contributed to these image collections.
\end{acks}

\bibliographystyle{ACM-Reference-Format}
\bibliography{refs}

\end{document}